\documentstyle[editedvolume,numreferences]{crckapb}

\begin{opening}
\title{ASTROPHYSICAL SOURCES OF GRAVITATIONAL RADIATION\protect\\
AND PROSPECTS FOR THEIR DETECTION}

\author{\'Eanna \'E. Flanagan}
\institute{Center for Radiophysics and Space Research\\
	Cornell University, Ithaca, NY 14853, USA.}

\end{opening}

\runningtitle{SOURCES OF GRAVITATIONAL RADIATION}

\begin{document}
\input psfig.tex

\def\agt{
\mathrel{\raise.3ex\hbox{$>$}\mkern-14mu\lower0.6ex\hbox{$\sim$}}
}
\def\alt{
\mathrel{\raise.3ex\hbox{$<$}\mkern-14mu\lower0.6ex\hbox{$\sim$}}
}
\def\beq{\begin{equation}}
\def\endeq{\end{equation}}

\begin{abstract}

\noindent
In the coming decade, the LIGO/VIRGO network of ground-based
kilometer-scale laser interferometer gravitational wave detectors will
open up a new Astronomical window on the Universe: gravitational waves
in the frequency band $10$ to $10^4$ Hz.  In addition, if the
proposed, 5 million kilometer long, space based interferometer LISA
flies, another window will be opened in the frequency band $10^{-4}$
to $1$ Hz.  We review the various possible sources that might be
detected in these frequency bands, and the information that might be
obtainable from observed sources.  Several key possible sources are
inspirals and coalescences of neutron-star neutron-star and/or
neutron-star black-hole binaries; inspirals, mergers, and
ringdowns of black-hole black-hole binaries (both solar mass and
supermassive); stellar core collapse; rapidly rotating neutron stars;
the formation of supermassive black holes; and inspirals of compact
objects into supermassive black holes.
\end{abstract}

\section{Introduction}

This review of gravitational wave sources is
divided into three
sections.  First, we review the detector sensitivities that
have been achieved to date and discuss projected
sensitivities for detectors now under construction.  Second, we summarize
the current observational upper limits on gravitational waves in
various frequency bands.

The main body of this review will consist of a survey of various
anticipated sources of waves.
Each anticipated source can be roughly characterized by a
characteristic frequency $f$ and a characteristic value of strain
amplitude $h$.  However, it is important to also note that sources
vary widely with respect to how uncertain is the rate of their
occurrence in the Universe.  The enterprise of anticipating potential
gravitational wave sources is extremely uncertain.  For most
sources that we can conceive of, either
the wave strengths are uncertain by several orders of magnitude, or
the event rate is uncertain by several orders of magnitude, or the
very existence of the source itself is very uncertain.  While there
are some important exceptions such as coalescing compact binaries,
these are the exception rather than the rule.  The upside of this
great uncertainty is the potential for gravitational wave  astronomy to
bring us new and interesting information.


For more details on the topics discussed here, the reader is
encouraged to consult the detailed recent review articles by Thorne
\cite{300yrs,thorne95,thorne97a,thorne97b}, and also the review
article by Allen \cite{allen96} on stochastic gravitational waves.

\section{Detector Sensitivities: Achieved and Projected}

As is well known, there are two technologies which are being pursued for
detection of gravitational waves.  The first is to monitor the modes
of vibration of a solid test mass which is cooled to very low
temperatures.  Such detectors are called resonant mass detectors, or
more colloquially ``bars''.  The second technology is to monitor the
relative displacement of
suspended test masses using laser interferometry.

There are presently operating several cylindrical resonant
mass antennae: the ALLEGRO detector at Louisiana State University
\cite{bar:allegro}, the NIOBE detector at the University of Perth,
Australia \cite{bar:niobe}, the EXPLORER detector at the University of
Rome \cite{bar:explorer}.  All of these detectors are operating at
temperatures of a few degrees Kelvin and have sensitivities of order
$h \sim {\rm (several)\ }  \times 10^{-19}$.  There are also one or
two so-called third generation detectors which are operating at
temperatures of a few tens of miliKelvin: the AURIGA detector at the
University of Legarno,
Italy \cite{bar:auriga}, and the NAUTILUS detector at the University of
Rome \cite{bar:nautilus}.  These detectors have sensitivities
of order $ h \sim {\rm (several)\ } \times 10^{-20}$ \cite{bar:nautilus}.


In addition, there are plans to construct a new generation of resonant
mass detectors which will be spherical, or nearly spherical (TIGAs or
Truncated Icosahedral Gravitational Arrays \cite{tiga}), or hollow
spheres.  Detectors in the planning stage include the GRAIL project in the
Netherlands \cite{grail}, the OMNI-1 project in Brazil, and the TIGA
project in the USA \cite{tiga}.  These detectors have design
sensitivities of order $10^{-21}$; see Fig.\ \ref{fig:summary} below.
Attaining this sensitivity level will require reducing the so-called noise
temperature by several orders of magnitude, and so will not be an easy
task.  More details on resonant
mass detectors can be found in the contribution by Massimo Cerdonio to
this proceedings.

Interferometer detectors have several advantages over resonant mass
detectors, the primary one being that they are intrinsically broadband
while resonant detectors are narrowband.  The technology for such
detectors has been continuously under development for several decades.
Today an international network of kilometer-scale, interferometer
detectors is under construction, consisting of:

\begin{itemize}
\begin{itemize}

\item  The American LIGO project \cite{ligoscience}, which initially
will consist of
three separate interferometers, an interferometer with $4 \, {\rm km}$
arms in Livingston, Louisiana, and two interferometers with armlengths
of $2 \, {\rm km}$ and $4 \, {\rm km}$ in Hanford, Washington.  The
detectors will come online around 2001.

\item  The French-Italian VIRGO project \cite{virgo}, which is
constructing an interferometer with $3 \, {\rm km}$ arms near Pisa,
Italy.  This detector is also planned to come online around 2001.

\item  The British-German GEO600 project \cite{geo}, which is
constructing a $600$ meter long interferometer near Hanover, Germany,
coming online around the year 2000.

\item The Japanese TAMA project \cite{tama}, which is constructing a
$300$ meter interferometer near Tokyo.

\end{itemize}
\end{itemize}

There are also several prototype interferometers, for instance, the
LIGO 40m prototype and the Garching and Glasgow interferometers
associated with the GEO 600 project.  The kilometer-scale
interferometers will act as a coordinated network: for each burst of
waves, the data from all the detectors will be analyzed to obtain the
two independent waveforms $h_+(t)$ and $h_\times(t)$, and also the
direction to the source.  More details on this network can be found in
the contribution by Norna Robertson to these proceedings.

Both resonant mass detectors and earth-based interferometer detectors
are limited to the frequency band $10 \, {\rm Hz} \alt f \alt 10^4 \,
{\rm Hz}$.  At lower frequencies, the background of time varying
near-zone Newtonian gravitational fields produced by objects on the
Earths surface (winds in the atmosphere, seismic waves in the Earth's
crust etc.) becomes very large.  In order to attempt to measure
gravitational waves in the band of frequencies $10^{-4} \, {\rm Hz}
\alt f \alt 1 \, {\rm Hz}$, it is necessary to go into space.  Laser
interferometry can be used to monitor the distances between spacecraft
and thus to measure gravitational waves.  One proposed mission of this
kind is the {\it Laser Interferometer Space Antenna} (LISA), which
the European Space Agency tentatively plans to fly around 2015
\cite{cornerstone}.  LISA consists of six spacecraft in a triangular
configuration in a solar orbit.

\begin{figure}
{\psfig{file=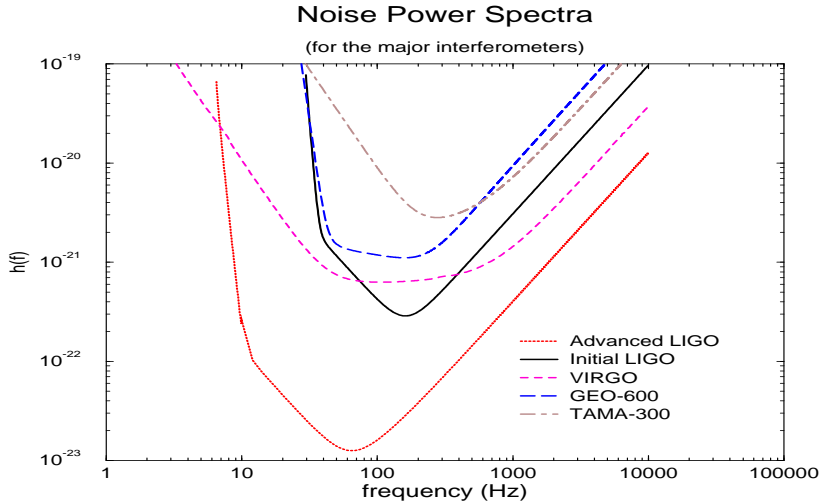,height=7cm,width=12cm,angle=-90}}
\caption{The projected noise spectra for the current major interferometer
projects.  The LIGO spectra were taken from Fig.10 of Ref.\
\protect\cite{ligoscience} and the spectra for the other
interferometers were obtained from Ref.\ \protect\cite{GRASP_data}.}
\label{fig:network}
\end{figure}

Figure \ref{fig:summary} shows the noise levels that have been achieved
to date in several representative detectors: bar detectors in
1976, in 1990 (Louisiana/Rome), in 1997, and the Caltech 40 meter
prototype interferometer in 1990 and 1994.
Also shown in Fig.\ \ref{fig:summary} are the projected noise levels for
future detectors: the first generation ``initial'' LIGO
interferometers, later ``advanced'' LIGO interferometers, the LISA
interferometer, and proposed TIGA resonant mass detectors.
There will also be a generation of interferometers in LIGO
intermediate between the initial interferometers and the advanced
interferometers; these ``enhanced interferometers'' will be obtained
by incrementally improving the various elements of the initial
interferometers over a time scale of a few years.
In Fig.\ \ref{fig:network} we compare the noise spectra for all the
major interferometer projects.

\begin{figure}
{\psfig{file=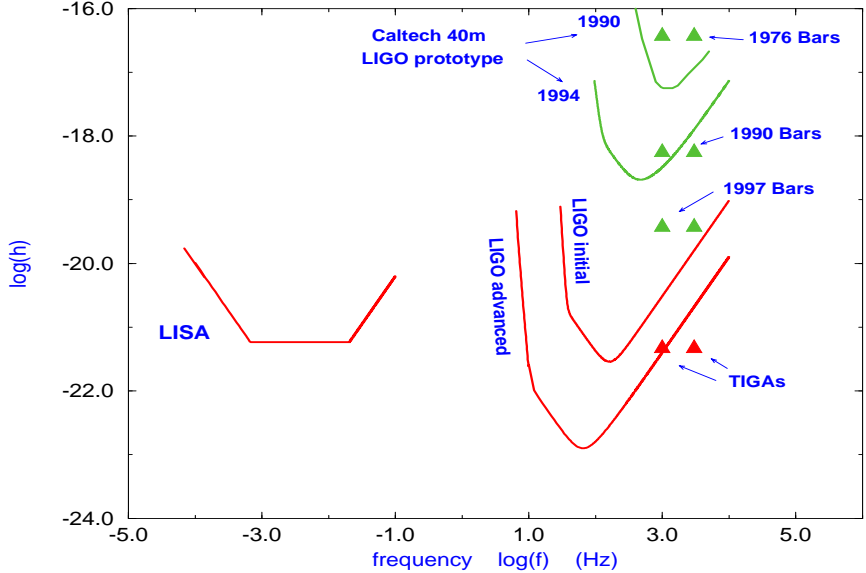,height=9cm,width=13cm,angle=-90}}
\caption{This plot shows the noise levels that have been achieved to date in
several gravitational wave detector systems: the Caltech 40m interferometer
\protect\cite{40meterdata}, and the best bar detectors in 1976,
\protect\cite{300yrs}, in the early 1990's \protect\cite{thorne95}, and in
1997 \protect\cite{bar:nautilus}.
It also shows that projected noise levels for future detector systems:
initial and advanced LIGO interferometers \protect\cite{ligoscience}, the
spacebased LISA interferometer \protect\cite{thorne95}, and future generations
of resonant mass detectors in the form of truncated icosahedral gravitational
arrays (TIGAs) \protect\cite{TIGAdetails}.
For the broadband interferometer detectors the quantity plotted is
$h_{\rm rms}(f) \equiv \protect\sqrt{f S_h(f)}$.  For the
narrowband resonant mass detectors (bars and TIGAs), the quantity
plotted is $h_{\rm rms,bar}  \equiv \alpha
 h_{\rm rms}(f_0) \protect\sqrt{f_0 / \Delta f}
$, where $\Delta f$ is
the bandwidth of the noise spectrum, $f_0$ is the central
frequency, and $\alpha$ is a constant of order unity
chosen so that narrowband and
broadband detectors will have the same
value of $h$ on this plot if they are equally efficient at detecting
broadband, burst waves \protect\cite{details}.
}
\label{fig:summary}
\end{figure}

For the broadband, interferometer detectors, the quantity plotted in
Figs.\ \ref{fig:summary} and \ref{fig:network} is the the
rms dimensionless strain per unit logarithmic frequency, $h_{\rm
rms}(f)$.  This is defined so that the total noise-induced
fluctuation in the output $h(t)$ of any detector can be written as
\begin{equation}
\langle h(t)^2 \rangle = \int_{-\infty}^\infty \, d (\ln f)  \ h_{\rm
rms}(f)^2.
\end{equation}
For the resonant mass detectors, on the other hand, the
quantity $1/h_{\rm rms}(f)$ sharply peaked, and the signal-to-noise
ratio for any broadband burst of waves is proportional to
$\int df 1 / h_{\rm rms}(f)^2 = \Delta f / h_{\rm rms}(f_0)^2$, where
$\Delta f$ is the bandwidth and $f_0$ is the central frequency.  In
Fig.\ \ref{fig:summary} we plot for resonant mass detectors the quantity
\beq
h_{\rm rms,bar} \equiv  \alpha \, \sqrt{f_0 \over \Delta f} \, h_{\rm
rms}(f_0),
\endeq
where $\alpha$ is a constant of order unity chosen so that narrowband
and  broadband detectors will have the same value of $h$ if they are
equally efficient at detecting broadband, burst waves \protect
\cite{details}.   [Note that, when considering the detection of
periodic signals, one should instead compare directly the values of $h_{\rm
rms}(f)$ of bars and interferometers.]

For a given broadband burst of waves, broadband detectors can
measure the gravitational waveform $h(t)$, whereas narrowband
detectors can essentially measure only the Fourier transform of the
waveform at the resonant frequency.  Although we have treated the planned
TIGA/spherical detectors as narrowband instruments above, in fact they
will have $\Delta f / f_0 \sim 0.1$ \cite{tiga}, in contrast to
bar detectors to date which have $\Delta f / f_0 \sim 10^{-3}$.  A
``xylophone'' of TIGAs could thus act as a broadband instrument \cite{tiga}.

Consider now how strong waves have to be in order to be detected.
There are three different types of gravitational waves: burst waves,
periodic waves and stochastic waves.  For each type of wave, let
$h_{\rm amp}(t)$ denote the physical size of the metric perturbation.
For burst waves, a signal with characteristic frequency $f_c$ will
be detectable whenever $h_{\rm eff} \agt h_{\rm rms}(f_c)$, where
the effective strain amplitude $h_{\rm eff}$ is
\beq
h_{\rm eff} = h_{\rm amp} {\sqrt{ {\cal N}_{\rm cycles}} \over \sqrt{ 2
\ln \left[ T/\Delta t \right] }}.
\label{heffburst}
\endeq
Here ${\cal N}_{\rm cycles}$ is the number of cycles of the waveform
in the bandwidth of the detector, $T$ is the total duration of the
data set in which one searches for a signal, and $\Delta t$ is the
effective sampling time \cite{noteburst}.  For periodic waves, a
signal with frequency
$f_c$ will be detectable when $h_{\rm eff} \agt h_{\rm rms}(f_c)$,
where
\beq
h_{\rm eff} = h_{\rm amp} {\sqrt{ T \Delta f } \over \sqrt{ 2
\ln \left[ T \Delta f \right] } }
\label{heffperiodic}
\endeq
and $\Delta f$ is the range of frequencies one searches through.
Finally, stochastic waves will be approximately detectable when
$h_{\rm eff} \agt h_{\rm rms}$, where
\beq
h_{\rm eff} = h_{\rm amp} \sqrt{ T \Delta f }
\label{heffstochastic}
\endeq
and $\Delta f$ is the bandwidth over which the detector is
sensitive.  Plotting $h_{\rm eff}$ versus $h_{\rm rms}$ for several
different detectors and various different sources of waves
illustrates the waves' detectability.  The definitions
(\ref{heffburst}) -- (\ref{heffstochastic}) are approximate but
correct to within factors of order unity; more precise
characterizations of detectability for burst, periodic and stochastic
waves can be found in Refs.\ \cite{300yrs}, \cite{cutlerperiodic} and
\cite{allen96}, respectively.

\section{Indirect detections and observational upper limits}

\begin{figure}
{\psfig{file=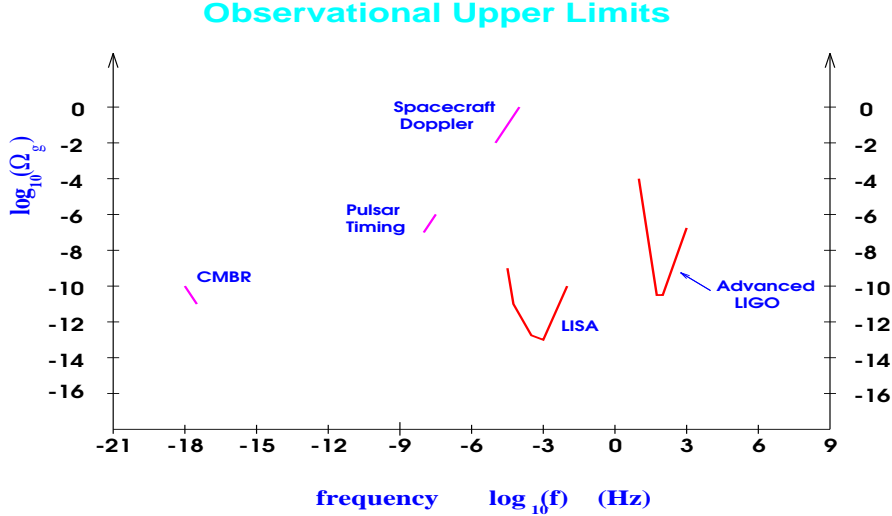,height=7cm,width=12cm,angle=-90}}
\caption{Current observational upper limits on gravitational wave
energy density in various frequency bands, together with the projected
sensitivities of advanced LIGO interferometers and of LISA.}
\label{fig:upperlimits}
\end{figure}

The most celebrated indirect detection of gravitational waves is of
course the monitoring of the orbital decay of the binary pulsar PSR
1913+16, for which the decay rate agrees with the prediction of
general relativity to better than $1 \, \%$ \cite{taylor}.
Aside from this, several different techniques have been used to place
upper limits on
the energy density in gravitational waves in different frequency
bands.   One technique is to
monitoring the radio signals from interplanetary spacecraft; such
signals would be Doppler shifted in a characteristic way by a burst of
gravitational waves.  The limits obtained from this technique
\cite{300yrs} are
illustrated in Fig.\ \ref{fig:upperlimits}.  Another technique is to
monitor precisely the arrival times of radio pulses from pulsars
(accuracy $\sim 1 \, \mu{\rm s}$) over several years; the resulting
limits \cite{pulsartimimg} are shown in Fig.\ \ref{fig:upperlimits}.
{}From primordial nucleosynthesis we obtain the constraint on
primordial gravitational waves that $\int d \ln f \, \Omega_g(f) \alt
10^{-5}$ \cite{allen96}; this limit does not apply to waves that were
produced after neucleosynthesis.
Finally, it is likely that some portion of the anisotropy of the cosmic
microwave
background radiation is due to primordial gravitational waves (the
remaining portion being due to density or scalar perturbations).  The
current observations of the anisotropy spectrum cannot distinguish
between scalar and gravitational wave contributions.  Thus, these
observations provide upper
bounds on $\Omega_g(f)$ at low frequencies as illustrated in Fig.\
\ref{fig:upperlimits} \cite{allen96}.  They {\it may} also constitute
an indirect detection of gravitational waves, although current
theoretical prejudice suggests that the gravitational wave
contribution to the anisotropy is smaller than the noise level in the
observations to date.

The limits in Fig.\ \ref{fig:upperlimits} are shown in terms of the
quantity $\Omega_g(f)$, which is the energy density
in gravitational waves per unit logarithmic frequency divided by the
critical energy density $\rho_{\rm critical}$ necessary to close the
Universe:
\begin{eqnarray}
{\Omega_g(f)} &\equiv& {1 \over \rho_{\rm critical}} \,
{d E \over d \ln f \, d^3 x}(f) \\
& = & \left( {f \over 100 \, {\rm Hz} } \right)^2 \, \left(
{h_{\rm amp} \over 1.3 \times 10^{-20} }\right)^2 \,
\left( {100 {\rm km} \, {\rm s}^{-1} \, {\rm Mpc}^{-1} \over H_0 }\right)^{-2},
\nonumber
\end{eqnarray}
where $H_0$ is the Hubble constant.

\section{Overview of Sources}

As discussed above, the the high frequency band of
$10 \, {\rm Hz} \alt f \alt 10^4 \, {\rm Hz}$ is the domain of
earth-based instruments.  Key sources in this band are stellar core
collapses, coalescences of compact binaries (neutron star -- neutron
star, neutron star -- black hole, and black hole -- black hole binaries),
vibrating/precessing neutron stars, and black hole births.
The low frequency band of
{$10^{-4} \, {\rm Hz} \alt f \alt 10^{-1} \,
{\rm Hz}$} is the domain of LISA.  Here the sources are
binary stars, formation and coalescences of supermassive black holes
(SMBHs), and capture of compact objects by SMBHs.
Sources of stochastic waves include parametric
amplification of vacuum fluctuations during inflation; defects such as
cosmic strings, and phase transitions in the early Universe.  These
sources can produce waves over a huge frequency band from {$f \sim
10^{-18} \, {\rm Hz}$} to {$f \sim {\rm MHz}$}.  In addition, the
superposition of very many indistinguishable coalescing compact binary
signals or stellar core collapse signals may produce a detectable
stochastic background in the high frequency band \cite{blair}.
Finally, more speculative sources include theoretical constructs such
as naked singularities, boson stars, soliton stars, etc.

\section{Stellar Core Collapse}

Type II supernovae occur roughly once every 40 years in our Galaxy.
The rate of stellar core collapses could be somewhat
larger than this, since some stellar core collapses could be
electromagnetically quiet and/or associated with the accretion induced
collapse of white dwarfs.  While the electromagnetic signal is
dominated by the ejected mantle, the gravitational wave signal is
dominated by the dynamics of the collapsing core.  Numerical modeling
of the dynamics of core collapse and bounce is complex: despite much
work in this area there is not yet a firm consensus.  The amount and
characteristics of the gravitational wave emission depends on
hydrodynamic processes that have not yet been well modeled, and also
on the initial rotation rate of the degenerate stellar core before
collapse, which is poorly known.  Thus, there are large uncertainties
associated with this type of source.  In fact, our knowledge is so
poor that the event detection rate could be as large as many per year
with initial LIGO interferometers, or as poor as less than one per
year with advanced LIGO interferometers.

There is a critical value of the initial angular momentum $J$ of the
degenerate stellar core.  If $J=0$,
then the core will bounce once nuclear densities are reached at $r
\sim 10 \ {\rm km}$, where $r$ is the radius.  Otherwise the
eccentricity $e$ increases during the
collapse; to a good approximation
\beq
{T \over W} \sim e^2 \sim {J^2 \over G M^3} \, {1 \over r},
\endeq
where $T$ is the rotational kinetic energy of the core, $W$ is its
gravitational potential energy, $J$ is its conserved angular momentum,
and $M$ is its mass.  The bounce
occurs at eccentricities of order unity when
$$
J \sim J_{\rm crit} \sim \sqrt{ {G M^3  \over 10 \, {\rm km}}},
$$
which corresponds to initial rotation periods of a few seconds.
For $J \gg J_{\rm crit}$, centrifugal flattening will hangup the
collapse.
On the other hand, if {$J \ll J_{\rm crit}$}, there is no
centrifugal hangup and it is thought that core remains axisymmetric
during the collapse.  Numerical simulations predict that in this case
the total energy radiated from the collapse and bounce
is {$\Delta E_{\rm rad}
\sim 10^{-7} M_\odot c^2$}, and the effective strain amplitude is
\cite{SNaxisymmetric}
$$
h_{\rm eff} \sim 3 \times 10^{-22} \left( {30 \, kpc \over r} \right).
$$
Such a core collapse would be visible only in the local group of
galaxies by LIGO/VIRGO/GEO.

Consider now the case in which $J \gg J_{\rm crit}$.  Here centrifugal
flattening and hangup occur before nuclear densities are reached; the
protoneutron star could form a disc with radius $\sim 100 \, {\rm km}$
or larger.  If $T/W \agt 0.3$, theory suggests that a the fluid
configuration is unstable to becoming non-axisymmetric via a dynamical
triaxial instability.  This would greatly enhance the gravitational
wave emission; a purely axisymmetric body can emit waves only via its
collapsing motion, whereas a non-axisymmetric body can emit waves via
its rotational motion.  Such a triaxial instability has been seen in
numerical simulations \cite{centrella}, in which the hydrodynamical
processes are treated in a simple and approximate way.
[Note however that one recent numerical simulation \cite{mueller97}
finds that the non-axisymmetry does not appreciably enhance the
gravitational wave emission contrary to expectations.] {\it If} such
an instability occurs, as much as $\sim 10^{-3} M_\odot c^2$ of energy
could be radiated into gravitational waves, and such collapses could
be seen out to the VIRGO cluster by initial LIGO interferometers and
to several hundred Mpc by advanced LIGO interferometers.
However, the fraction of core collapses (if any) which become very
non-axisymmetric is unknown.
Nevertheless, even if only one core collapse in
$10^4$ undergoes a tri-axial instability, such collapses might be the
most common type of core collapse seen by the detectors.  Several
specific scenarios for non-axisymmetry are reviewed by Thorne
\cite{thorne97b}.

There is little observational evidence concerning the
initial rotation rates of degenerate stellar cores.  Two points are
worth noting, however.  First, the increasing
evidence that neutron stars undergo substantial kicks in supernovae
\cite{chernoff} (achieving velocities $\agt 1500 \, {\rm km} {\rm
s}^{-1}$ in some cases)
suggests that core collapses are very non-symmetric.
Second, observations suggest that newborn neutron stars are
typically {\it not} born spinning near the breakup angular velocity
that one would expect if the core collapse has an excess of angular
momentum \cite{reviewNS}.  The traditional view has been that this
indicates that most
pre-collapse stellar cores are slowly rotating.  However, it has
recently been
predicted that newborn neutron stars can spin down from near maximal
rotation down to $\sim 20 \ {\rm Hz}$ within a year or so due to
$r$-mode instabilities (see Sec. \ref{sec:periodic} below).  Thus,
there need not be a contradiction between the observational evidence
that rapidly spinning (periods $\sim $ milliseconds) neutron stars have
typically accreted most of their angular momentum, and the supposition
that most pre-collapse degenerate stellar cores could be rapidly
rotating.

Detecting stellar core collapses would have several payoffs for
astronomy \cite{300yrs}.
If the core collapse were close enough to produce a detectable flux of
neutrinos, one could constrain neutrino masses by comparing the
arrival times of gravitational waves and neutrinos.  For example, a
coincident detection of gravitational waves and neutrinos from the
supernova SN1987a would have yielded the limit $m_\nu \alt 4 \, {\rm
eV}$.  One could also obtain information from the gravitational
waveforms about the dynamics of the core collapse and bounce, tipoff
optical astronomers before optical supernova first brighten (so they
can measure the early portion of the light curve), and obtain
observational evidence about the rate of formation of neutron stars.

\section{Coalescences of Compact Binaries}

\subsection{Neutron star -- Neutron star coalescences}

The most reliable source for the LIGO/VIRGO/GEO network is the
coalescences of neutron star/neutron star (NS/NS) binaries; this type
of source has been studied theoretically in great detail over the last
several years.  The event rate is fairly well understood from
observations of progenitor NS/NS systems in our own Galaxy: the
distance to which one must look to see 3 events per year has been
estimated to be $\sim 200 \, {\rm Mpc}$ \cite{phinney,narayan}.
[Recent revisions to the pulsar distance scale indicate that this
distance could be somewhat larger, perhaps $\sim 400 \, {\rm Mpc}$
\cite{taylor98}].  The range of initial LIGO interferometers for NS/NS
binaries is $\sim
25 \, {\rm Mpc}$, while that of the enhanced interferometers will be
$\sim 250 \, {\rm Mpc}$.   There is a clean separation in the
gravitational waveforms between an {\it inspiral} phase of the signal,
which carries most of
the detection signal-to-noise ratio and which is fairly well
understood theoretically, and a subsequent {\it merger} phase, which
depends on the internal structure of the stars and which is poorly
understood theoretically \cite{details1}.  Fortunately,
understanding the details of the merger phase is not necessary for
detecting the signals; for detection one will use the inspiral waves.

Detection of gravitational waves from NS/NS binaries would test the
fairly popular theory that NS/NS mergers are the source of gamma ray
bursts, which are seen at a rate of roughly once per day by satellite
gamma ray detectors, and whose origin has been a longstanding puzzle.
The rate of burst detections of a few per day is roughly consistent
with the inferred rate of NS/NS mergers out to cosmological distances,
and amount of energy supplied by a NS/NS merger is (marginally)
sufficient to produce the observed $\gamma$ ray energy fluxes for a
source at a cosmological distance.  In the last year new evidence has been
uncovered that gamma ray bursts are cosmological in origin rather than
Galactic, which favors the NS/NS coalescence hypothesis: optical
counterparts of some particular bursts have been seen at cosmological
distances \cite{metzger}.
On the other hand, while the light
curves and spectra of the bursts seem to be well fit by a model of an
expanding relativistic fireball, there are theoretical difficulties in
understanding how NS/NS mergers (or NS/BH or BH/BH mergers) could give
rise to such fireballs with sufficiently low baryon contamination.
[Too many baryons in the fireball would convert too much of the
fireball's energy into kinetic energy of the baryons instead of into
radiation].


During the inspiral of the two stars, general relativistic effects are
large as the orbital velocity is about $\sim 20 \%$ of the speed of
light at the point where most of the detection signal-to-noise ratio
is obtained.  The signal is very clean however: to a very good
approximation, the two stars can be treated as point masses with
intrinsic spins.  The gravitational waveform is then determined by the
relativistic gravitational interaction between the two point masses;
all of the other complicated physical effects that occur
(heating and tidal distortion of the stars, interactions involving the
stars' magnetic fields etc.) have a negligible effect on the
gravitational waveforms in the inspiral phase \cite{bildstencutler}.

This cleanliness means that the characteristics of the waveforms can
be predicted in advance, and that the technique of matched filtering
\cite{300yrs} can be used to detect the waves.  By cross-correlating the
outputs of the various interferometers in the network with theoretical
template waveforms, one can both detect the waves and also measure
the binaries parameters: masses, spins, orbital elements, sky
location, distance etc.  Very high accuracy templates are not required
to detect the waves: currently available, post-2-Newtonian templates
will be adequate in most cases \cite{poisson97}.  By contrast, in
order to extract the parameters of the binary with acceptably small
systematic errors, very high accuracy templates will be required; such
templates are under construction \cite{blanchet1}.  It is possible
that approximate 
post-Newtonian waveforms may be substantially improved using the method
of Pad\'e approximants \cite{damour97}.  Much theoretical work has
been done on how
accurately the parameters of the binary can be measured; however, the
parameter-extraction accuracies are currently only known to within
factors of order 2.

There are several payoffs of measuring the inspiral waves from
coalescing binaries.  One will be able to experimentally measure for
the first time various non-linear aspects of general relativity such
as gravitational wave tails \cite{cutlerflanagan94,blanchet2} and
possibly in some cases the 
dragging of inertial frames \cite{last3minutes}.  It will be possible
to constrain non general-relativistic theories of gravity, for example
one will be able to improve on solar system bounds on the
dimensionless parameter $\omega$ of Brans-Dicke theory \cite{will94},
and one will obtain constraints on the ``graviton mass'' \cite{will98}.
One could identify the sources of gamma ray bursts by
comparing their arrival times to those of gravitational wave bursts.
If the event rate is large enough, one will be able to accumulate
statistics on the distances, approximate sky locations, and masses of
many ($\sim 100$) coalescences.  This information will directly probe
the distribution of Galaxies on scales $400 \, {\rm Mpc} \alt D \alt 4
\, {\rm Gpc}$ with a resolution of $\sim 200 \, {\rm Mpc}$
\cite{cutlerflanagan94}; the
distribution of Galaxies on such large scales has not yet been probed
in redshift surveys.  In addition, one will be able to attempt to
measure the three cosmological parameters: the Hubble constant, the
cosmological constant, and the closure parameter $\Omega$.  Monte
Carlo simulations suggest that with one year's
observations with advanced LIGO interferometers, one will be able to
measure the Hubble constant to $\sim \, 1 \%$ and the other two
parameters with fractional accuracies of a few tens of percent
\cite{markovic94}.  [Of course, it is likely that these parameters
will already have been accurately measured by the time advanced
interferometer
sensitivity levels are achieved, by the MAP and PLANCK satellite
measurements of the CMBR \cite{CMBRref}].

Turn, now, to the gravitational waves from the merger stage of
coalescing binaries.  These will carry information
about the internal structure of the neutron stars and potentially about
the equation of state of nuclear matter at high densities.   Efforts
are underway to understand the dynamics of the merger process and the
features of the emitted waves using numerical simulations.  So far,
simulations of the merger have been limited to the Newtonian or
post-1-Newtonian approximations \cite{centrella1}.  There are
indications that it might be necessary to simulate the merger
numerically in full general relativity; efforts are underway in this
direction.
The merger waveforms will have most of their power at high frequencies
of several hundred Hz or higher.  Unfortunately, broadband laser
interferometers have poor performance at these high frequencies (see
Fig. \ref{fig:summary} above).  However, it might be possible to
measure the merger waves with specialized narrowband interferometers
which are designed to have enhanced sensitivity near some chosen
frequency and worsened sensitivities elsewhere \cite{narrowb}, or with
the next generation of resonant mass antennae (TIGAs or spheres).

\subsection{NS/BH and BH/BH binaries}

\begin{figure}
{\psfig{file=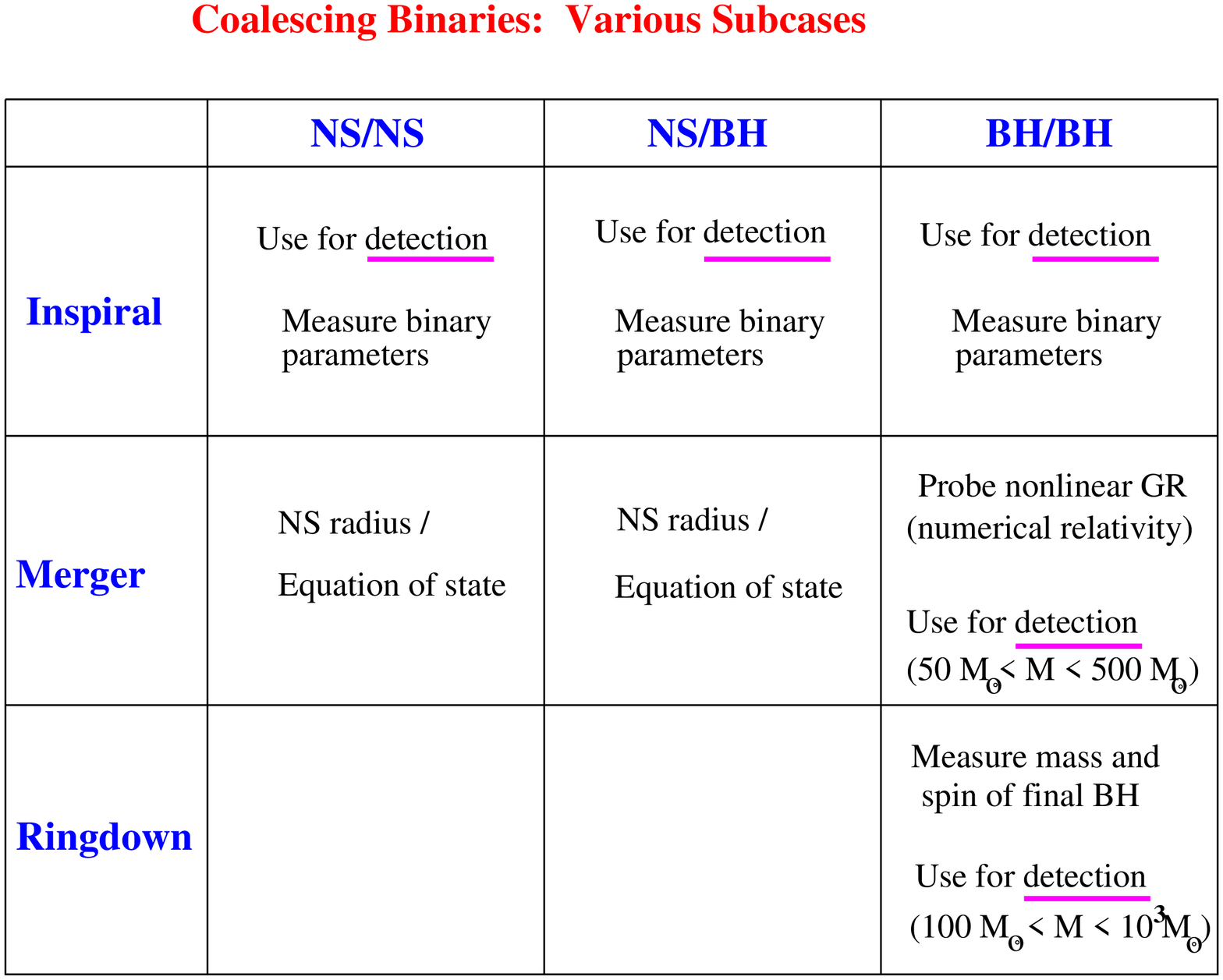,height=8cm,width=12cm,angle=-90}}
\caption{A summary of the various phases of the waves (inspiral,
merger, and ringdown) for the three different types of coalescing
binary, and the usefulness of each phase for each type of source.}
\label{fig:otherbinaries}
\end{figure}

The other types of coalescing compact binary that might be measured
are NS/BH and BH/BH binaries, as illustrated in Fig.\
\ref{fig:otherbinaries}.  For the NS/BH case the signal can be divided
into inspiral and merger phases as before; for the BH/BH case the
signal can be divided into three phases, inspiral, merger and ringdown
\cite{flanaganhughes}.

There are several fundamental differences
between the NS/BH and BH/BH cases, and the NS/NS case:

\begin{figure}
{\psfig{file=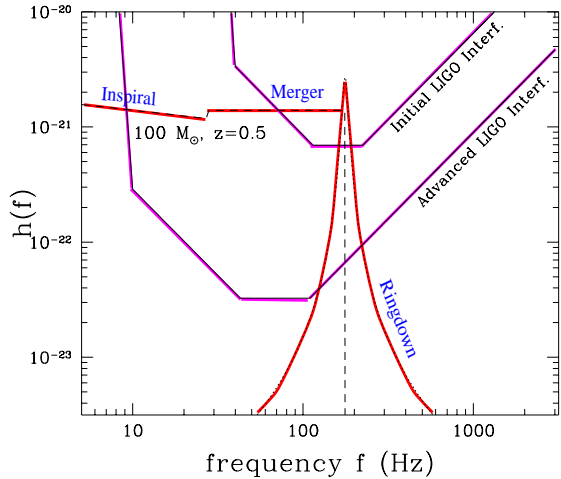,height=8cm,width=8cm,angle=-90}}
{\vskip -1.0cm}
\caption{A plot of the wave strength $\protect\sqrt{5} h_{\rm eff}$
for waves
from a binary black hole
coalescence of total mass $100 M_\odot$ at redshift $z=1$, together
with the noise levels $\protect\sqrt{5} h_{\rm rms}$ for initial and advanced
LIGO interferometers (adapted from Ref.\
\protect\cite{flanaganhughes}).  The energy spectra for the merger
waves and the
ringdown waves
are taken from the crude models of Ref.\ \protect\cite{flanaganhughes}
($10\%$ of total mass energy radiated in merger waves and $3\%$ in
ringdown waves). Notice that the inspiral waves are not visible with
the initial LIGO interferometers, and barely visible with the
advanced.  For this type of source, it will be necessary to try to
detect the waves using the merger or ringdown waves directly.}
\label{fig:BBHexample}
\end{figure}

First, the event rates for NS/BH and BH/BH coalescences are far less
certain than NS/NS coalescences, since so far we have no observed
progenitor systems.  Arguments based on progenitor
evolution scenarios suggest an event rate rate very roughly $\sim 3
\, {\rm yr}^{-1}$ within $200 \, {\rm Mpc}$, i.e., the same rate as
for NS/NS binaries, in both the NB/BH and BH/BH cases \cite{phinney}.
[A recent estimate by Brown and Bethe reduces this distance
to $100 \, {\rm Mpc}$ for the NS/BH case \cite{bethe}].  There are
three classes of BH/BH binaries:
field binaries, binaries formed by capture
processes in galactic centers, and capture binaries in globular
clusters.  Siggurdson and Hernquist have argued that there is generically at
least one BH/BH event per core collapse globular cluster \cite{sh};
this yields $\sim 3 $ BH/BH events per year within a distance of $600 \,
{\rm Mpc}$ using the extrapolation method of Sec.\ 3.1 of Ref.\
\cite{phinney}.  For field BH/BH binaries, recent estimates of the
coalescence rate by experts in binary evolution theory range from
$\sim 10^{-8} \, {\rm yr}^{-1}$ to $\sim 10^{-6} \, {\rm yr}^{-1}$ in
our Galaxy \cite{tutukov,sternberg}, to completely negligible
\cite{yungelson}.  There are large uncertainties associated with these
theoretical estimates of the coalescence rates.

Second, for BH/BH coalescences, the merger waves will bring
information about the dynamics of general relativity in a highly
dynamical, highly nonlinear, highly non-spherical regime that has
never before been probed experimentally.  A major effort is underway in
the numerical relativity community to simulate the mergers of black
holes, driven in
part by the possibility of performing comparisons with gravitational
wave data; see the article by Ed Seidel in this proceedings.  The
overall energy and the amount of information carried by the merger waves
are still very uncertain.

Third, the inspiral waves are stronger for NS/BH and BH/BH sources
than for NS/NS sources,
simply because of their greater mass.  The initial LIGO
interferometers can reach to $\sim 40 \, {\rm Mpc}$ for a BH/BH binary of
total mass $M = 5 \, M_\odot$ and to $\sim 300 \, {\rm Mpc}$ for $M = 50
M_\odot$ \cite{flanaganhughes}; enhanced interferometers can see
roughly ten times further and advanced interferometers about 30 times
further.  Thus, BH/BH coalescences might be detected early by the
LIGO/VIRGO network, and may be seen before NS/NS coalescences.  Note
also that if BH/BH events are seen, the most commonly detected events
will be the largest mass systems, since the increase in
signal-to-noise ratio with increasing mass very likely more than
compensates for the decrease in the mass function with increasing mass
\cite{KT}.

Fourth, the frequency at which the inpiral ends, i.e., at which the
spinning-point-mass approximation fails, decreases as the total mass
of the system is increased.  For sufficiently massive systems, the
regime of optimum sensitivity for LIGO ($f \sim 100 \, {\rm Hz}$)
overlaps with the complicated merger waves and not with the simple
inspiral waves; see Fig.\ \ref{fig:BBHexample}.  For such systems one
needs to detect the events using the merger or ringdown waves; this
implies an extra importance for numerical relativity simulations
\cite{flanaganhughes}.  For black hole binaries
of total mass $\sim 20 M_\odot$ say (a conventional value), events can
be detected using inspiral waves; however a large part of the signal-to-noise
ratio is accumulated in a very relativistic regime $6 M \alt r \alt 12
M$ in which post-Newtonian detection templates are possibly
inadequate.  New numerical techniques are under development for
calculating waveforms in this so-called ``Intermediate Binary Black
Hole'' (IBBH) regime \cite{IBBH}.

\section{Periodic sources}
\label{sec:periodic}

As is the case for supernovae, the expected event detection rate for
periodic sources is highly uncertain due to our ignorance as to the
wave's sources.  The primary expected type of source are spinning
neutron stars.  Such stars can radiate in several ways \cite{schutz98}:

First, if the
axis of rotation is displaced from the corresponding principal axis of
the moment of
inertia tensor by a small ``wobble angle'' $\theta_w$, then the star
radiates at the precessional sideband of the rotation frequency with
amplitude proportional to $\theta_w \epsilon_p$, where $\epsilon_p =
(I_{xx} - I_{zz})/I_{zz}$ is the poloidal eccentricity \cite{ZS}.  (We
assume the axis of rotation is the $z$ axis).  However, the
wobble angle decreases exponentially due to gravitational wave
emission on the short timescale \cite{horvath94}
\begin{equation}
\tau_{\rm brake} \sim 0.03 \, {\rm yr} \left( \varepsilon_p \over 0.01
\right)^{-2} \left( f_{\rm rot} \over 100 \, {\rm Hz} \right)^{-4},
\endeq
where $f_{\rm rot}$ is the rotational frequency.  [See also Ref.\
\cite{wasserman} for another dissipative mechanism which could also
quickly damp away precessional motion.]
Thus, this mechanism for producing waves is not likely to be detected,
unless possibly if the neutron star is accreting and if the accretion
continuously drives $\theta_w$ away from zero, as has been suggested
by Schutz \cite{schutz98}.

Second, it is possible for the solid crust of the neutron star to
support some non-axisymmetry.   If the star is not axisymmetric,
it will radiate at twice the rotation frequency with an amplitude
proportional to the
equatorial ellipticity $\varepsilon_e = (I_{xx} - I_{yy}) / I_{zz}$
\cite{ZS}.  The effective strength of the waves is given by
\begin{equation}
h_{\rm eff} \sim 4 \times 10^{-20} \left({f_{\rm rot} \over 500 \,
{\rm Hz}}\right)^{5/2} \, \left( {T_{\rm obs} \over 1/3 \, {\rm yr}}
\right)^{1/2} \left( {1 \, {\rm kpc} \over r} \right)
\left( {\varepsilon_e \over 10^{-6} } \right).
\end{equation}
Here $T_{\rm obs}$ is the observation time and $r$ is the distance to
the source.  For advanced LIGO interferometers it follows that the
minimum detectable eccentricity is of order
\begin{equation}
\varepsilon_e \sim 10^{-7} \,\, \left({f_{\rm rot} \over 500 \, {\rm
Hz}}\right)^{-1} \,\, \left( {T_{\rm obs} \over 1/3 \, {\rm yr}}
\right)^{-1/2} \,\, \left( {r \over 1 \, {\rm kpc}} \right).
\end{equation}
With specialized narrowband interferometers \cite{narrowb} or resonant
mass detectors \cite{tiga}, this lower bound might be improved by two
orders of magnitude \cite{thorne97b}.

We are very ignorant as to the likely size of $\varepsilon_e$ in
spinning neutron stars.  Theoretical estimates of the breaking strain
of the star's crust yield the upper limit $\varepsilon_e \alt
10^{-5}$ \cite{ShapiroTeukolsky}.  The observed spin down rate of
pulsars sets an upper limit
on the eccentricity (obtained by assuming the entire spin down is due
to gravitational wave emission); this limit is of order $10^{-8}$ for
millisecond pulsars \cite{300yrs}.  However, it does not follow that
$\varepsilon_e$ should be this small for other populations of neutron
stars, whose typical evolutionary histories would be expected to be
very different from those of millisecond pulsars.  Several different
plausible mechanisms for generating equatorial eccentricity have been
suggested, for example the anisotropic pressure caused by magnetic
fields \cite{thorne95,bonnazola2}.

In the case of accreting systems (low mass X ray binaries), Bildsten
has recently
suggested that temperature gradients due
to non-uniform accretion over the surface of the star lead (via
temperature-dependent nuclear reactions) to density variations in the
crust and thence to a non-zero $\varepsilon_e$; the resulting
estimated values of $\varepsilon_e$ are of order $10^{-7}$
\cite{bildsten98}.   Bildsten also suggested an explanation for the
observational puzzle that the rotation rate of several observed such
systems
cluster around $300 \, {\rm Hz}$: gravitational wave emission is
preventing these sources from being spun up any
further, i.e., all the angular momentum being accreted is being
radiated into gravitational waves.  If this turns out to be true,
several known systems such as SCO-X1 would be detectable by advanced
LIGO interferometers \cite{bildsten98}.

A third mechanism by which spinning neutron stars can emit
gravitational waves is via the existence of unstable modes of
vibration of the star.  The $f$-mode of neutron stars will be unstable
to a radiation-reaction driven instability called the
Chandrasekhar-Friedman-Schutz instability at sufficiently rapid
rotation rates ($\agt 0.9 \Omega_{\rm breakup}$).  It has been
shown that this instability is counteracted  by the dissipative effects
of viscosity, except when the star's temperature lies in the narrow
band $10^9 \,K \alt
T \alt 10^{10} K$ \cite{lindblom}.  Neutron stars are this hot only in
the first few years after their formation, and newly born stars are
typically not likely to be born spinning rapidly enough for the
instability to operate.  Thus, $f$-mode, radiation-reaction driven
instabilities are not likely to be detected, except possibly in hot,
accreting systems \cite{schutz97}.  In addition, there are
viscosity-driven ``bar mode'' instabilities
which occur only for certain equations of state and for very rapidly
rotating ($P \sim 1 {\rm ms}$) stars \cite{bonnazola2}; again, such
instabilities are unlikely to be observable for the same reasons as for
the CFS instability.

Recently Anderson \cite{anderson97}, and also Friedman and Morsink
\cite{friedman97}, discovered that neutron star $r$ modes suffer from
an analogous radiation-reaction-driven instability.
Calculations indicate that the lowest $r$ mode is unstable
for temperatures $T$ in the roughly the same range as above, when
$\Omega \alt 20 \, {\rm Hz}$ or so \cite{lindblom98}.
Thus, newly born neutron stars should be
copious emitters of gravitational waves, and that they should always
spin down from their initial rotation rate to $\sim 20 \, {\rm Hz}$ on
the timescale of a year after their formation \cite{lindblom98}.  Such
strong sources
should likely be visible to the VIRGO cluster of galaxies with initial
LIGO interferometers, yielding an event rate of several per year.
It is important to firm up the initial, crude estimates of wave
strengths from these sources to verify this scenario.

Finally, one should note that searching for neutron stars or any other
type of periodic source is enormously computationally demanding due to
the necessity of correcting for Doppler shifts due to the Earths
motion and rotation, for many, many individual directions on the sky
\cite{cutlerperiodic}.  All sky searches will be {\it computation limited}:
the minimum detectable signal strength with a Teraflop computer will
be a factor of a few times larger than the ideal minimum detectable
signal strength one would obtain with infinite computing power.
Efforts are underway to devise sophisticated search algorithms to
minimize the loss in detectable signal strength
\cite{cutlerperiodic,schutz98}.

\section{Low Frequency Sources: Supermassive Black Holes}

Consider now the low frequency band $10^{-4} \, {\rm Hz} \alt f \alt
0.1 \, {\rm Hz}$, the domain of the proposed spacebased interferometer
LISA.  There is a guaranteed class of sources in this frequency band:
binary stars, including common envelope binaries, close white dwarf
binaries, contact binaries, cataclysmic variables, etc.  For many
binary, one should be able to combine optical and gravitational wave
information to solve for all the orbital elements of each binary and
to test general relativity.  At low frequencies, there will be so many
white dwarf binaries that they will be experimentally
indistinguishable and will constitute a source of background noise.  Also,
the shortest period NS/NS binary in the Galaxy should have a period of
about $\sim 10^5 \, {\rm yrs}$, from the estimates of NS/NS
coalescence rates.  Such a binary would be visible to LISA with a
signal-to-noise ratio of $\sim 500$ \cite{thorne95}.

LISA should also be able to see the formations and/or coalescences of
supermassive black holes (SMBHs) throughout the observable Universe with
signal to noise ratios of hundreds or thousands.
There is some observational evidence for
SMBH binaries: wiggles in the radio jet of QSO 1928+738 have been
attributed to the orbital motion of a SMBH binary
\cite{Roos}, as have time variations in quasar luminosities
\cite{sillanpana} and in emission line redshifts \cite{gaskell}.  The
overall event rate is uncertain, but could be large ($\agt 1/{\rm
yr}$), especially if the hierarchical scenario for structure formation
is correct \cite{Haehnelt}.  Detecting coalescences of supermassive
black holes would allow high precision tests of general relativity.

{}From the waves from a small black hole or compact object spiraling
into a larger black
hole, one can in principle obtain detailed information about the
spacetime geometry of the larger black hole (its multipole moments
\cite{ryanmap}), and test the no-hair theorem of general relativity
\cite{thorne97b}.  As a foundation for extracting this information,
theorists must understand how to calculate the motion of a test
particle in a Kerr geometry under the influence of gravitational
radiation reaction.  Some progress has been made recently in this
direction: a general expression for the radiation reaction force on a
test particle in any background spacetime has been derived by three
independent methods, all of which give the same answer
\cite{waldquinn}.  Ori \cite{ori} has proposed, without proof, a
practical calculational scheme to calculate the evolution of orbits:
it would be useful to verify that Ori's scheme can be derived from the
general Mino-Quinn-Wald equation of motion \cite{waldquinn}.

\section{Sources of Stochastic Gravitational Waves}

I will not attempt to give a review of stochastic sources
here, but instead refer the reader to the recent detailed and
comprehensive review article by Allen \cite{allen96}.

\section{Conclusion}

With the kilometer scale interferometer network about to come online
in the next few years, this is an exciting time for gravitational wave
astronomy.  Almost certainly we have not anticipated all the
detectable sources of gravitational waves that Nature produces in the
real Universe.  Hopefully she will bring us some surprises.

\section*{Acknowledgments}
I wish to thank the conference organizers for putting together such a
wonderful meeting.  The financial support of NSF grant PHY-9722189 is
also gratefully acknowledged.


\begin{thebibliography}{100}
\bibitem{TIGAdetails}
To obtain the value $h_{\rm rms,bar} \sim 5 \times 10^{-22}$ for TIGAs
in Fig.\ \ref{fig:summary} we used
\cite{details} with the parameters for one possible example of a
detector given in Ref.\ \cite{tiga}: mass $m = 2.5 \times 10^4 \,
{\rm kg}$ and resonant frequency $f_0 = 10^3 \, {\rm Hz}$.  We assumed
that the detector operates at the quantum limit $k T_n = h f_0$ and
that the total cross section $\int df \sigma_0(f) \sim 4 \times
10^{-20} \, cm^2 \, {\rm Hz}$ from Eq.\ (83) of Ref.\ \cite{300yrs}
which is approximately applicable to spherical detectors \cite{tiga}.
\bibitem{noteburst}
This is the standard relation that applies to matched filtering
\protect\cite{300yrs}; however it also applies to within factors of
$2$ or $3$ in situations where the waveforms shape is not well known
and matched filtering cannot be carried out
\protect\cite{flanaganhughes}.
\end{thebibliography}
\end{document}